\documentclass[10pt,conference]{IEEEtran}

\usepackage{epsfig}
\usepackage{lipsum,graphicx,subcaption}
\usepackage{multicol}
\usepackage{amsmath,amsthm,amssymb,amsfonts}
\usepackage{amssymb}
\usepackage{bm}
\usepackage[noend]{algpseudocode}
\usepackage{algorithmicx,algorithm}

\makeatletter

\newcommand{\Rmnum}[1]{\expandafter\@slowromancap\romannumeral #1@}
\makeatother
\begin{document}\sloppy
\def\x{{\mathbf x}}
\def\L{{\cal L}}

\pagestyle{empty}

\title{Privacy-preserving Sensory Data Recovery}

\author{Cai Chen$^{1}$, Manyuan Zhang$^{1}$, Huanzhi Zhang$^{1}$, Zhenyun Huang$^{1}$, Yong Li$^2$ \\ 
$^1$The Center for Information Geoscience, University of Electronic Science and Technology of China\\
$^2$School of Geophysics, Chengdu University of
Technology
}

\maketitle
%%%%%%%%%%%%%%%%%%%%%%%%%%%%%%%%%%%%%
\begin{abstract}
\thispagestyle{empty}
In recent years, a large scale of various wireless sensor networks have been deployed for basic scientific works. Massive data loss is so common that there is a great demand for data recovery. While data recovery methods fulfil the requirement of accuracy, the potential privacy leakage caused by them concerns us a lot. Thus the major challenge of sensory data recovery is the issue of effective privacy preservation. Existing algorithms can either accomplish accurate data recovery or solve privacy issue, yet no single design is able to address these two problems simultaneously. Therefore in this paper, we propose a novel approach \emph{Privacy-Preserving Compressive Sensing with Multi-Attribute Assistance} (PPCS-MAA). It applies PPCS scheme to sensory data recovery, which can effectively encrypts sensory data without decreasing accuracy, because it maintains the homomorphic obfuscation property for compressive sensing. In addition, multiple environmental attributes from sensory datasets usually have strong correlation so that we design a \emph{Multi-Attribute Assistance} (MAA) component to leverage this feature for better recovery accuracy. Combining PPCS with MAA, the novel recovery scheme can provide reliable privacy with high accuracy.

Firstly, based on two real datasets, IntelLab and GreenOrbs, we reveal the inherited low-rank features as the ground truth and find such multi-attribute correlation. Secondly, we develop a PPCS-MAA algorithm to preserve privacy and optimize the recovery accuracy. Thirdly, the results of real data-driven simulations show that the algorithm outperforms the existing solutions.
\end{abstract}
\begin{keywords}
 Data recovery, privacy-preserving, multi-attribute, wireless sensor networks
\end{keywords}
%%%%%%%%%%%%%%%%%%%%%%%%%%%%%%%%%%%%%%
\section{Introduction}
\label{sec:intro}
With the rapid development of Internet of Things (IoT) \cite{gubbi2013internet}, its applications become more and more widespread. There is a great demand for big data. We target at an important scenario that wireless sensor networks (WSNs) \cite{yang2014wireless}, as basic equipment, have been widely deployed to gather various sensory attributes from cooperative sensor nodes. These big sensor data have promising applications extensively adopted in military, civilian, medical treatment and commercial field. The origin data from sensor nodes are usually incomplete, which need to be recover. However, the recovery needs computing resources thus the missing data to process are transferred to the sink nodes or the server, during which the data packet may leak out. As some data exposes users' daily activities or confidential information, the privacy issue is a major concern in data recovery. Thus the major challenge of sensory data recovery is the issue of effective privacy preservation. In the practical application scenario, such as intrusion detection in battlefields, searching and rescuing systems, item tracking, infrastructure and environment monitoring, and indoor localization, valuable data demand not only accurate recovery but also effective protection against being stolen and stalked. Therefore we are motivated to design an efficient method for recovering incomplete data with privacy preservation.

For massive missing data recovery, there are plenty of classic missing data estimation methods such as KNN \cite{adeniyi2016automated}, DT \cite{hong2015reversible}, etc. These methods avoid the data leakage by requiring no data exchange. However, their recovery accuracy is usually limited. Currently, compressive sensing (CS) \cite{liu2017ls} is an advanced recovery technique with promising performance of effective estimation in mature applications. Yet, its data transmission and requirement for a computing server degrade privacy. For privacy, existing works are dummification and obfuscation. Although the two procedures protect privacy to some extent, they pollute the original data, decreasing the recovery accuracy. To improve the privacy, recently, another privacy method is presented in the trajectory field, namely K-vector perturbation (KVP) \cite{kong2015privacy}. The main idea of KVP is to use a private key to perturb a user's trajectory with K other trajectories while maintaining the homomorphic obfuscation property. Combining with compressive sensing, this novel encryption approach applied in the trajectory recovery is called privacy-preserving compressive sensing (PPCS) \cite{kong2015privacy}, which simultaneously tackles the challenges of privacy-preserving and recovery accuracy. 

In addition, a large quantity of real-world sensor data share linear correlation among multiple attributes, which can be used as the supplement of the internal correlations and improve the accuracy of the estimation. This is the so-called $\emph{Multi-Attributes Assistance}$ (MAA) method. In commercial WSNs, data from some kinds of expensive sensors (e.g PM2.5, CO2) are valuable and have something to do with data trading, but may cost too much for processing, like using PPCS for privacy. Thus with MAA, we can also bring benefits to cost control or data trading. Motivated by the these observations, we propose the novel PPCS with MAA approach which conquers all the mentioned challenges.

Our paper is organized as follows. Section II presents details about privacy-preseving recovery model and MAA statement with real-world data verification. Section III describes the procedures of our approach and gives an algorithm for encryption, multi-attribute assistant data recovery and decryption. Section IV introduces an evaluation of PPCS-MAA and implements a real-world data-driven experiment. Conclusions are made in Section V.

% %%%%%%%%%%%%%%%%%%%%%%%%%%%%%%%%%%%%%%%
\section{System Model and Problem Formulation/Statement}
In this section, we introduce the basic notations of the mathematical representation, privacy-preserving recovery model, multi-attribute assistance model, and the formal definition of our problem.
\subsection{Notations}
In this paper, scalars are denoted by lowercase letters, e.g., $n$; 
vectors are denoted by boldface lowercase letters, e.g., $\bm{x}$ and the transpose is denoted as $\bm{x}^{\top}$; matrices are denoted by boldface capital letters, e.g., $\bm{X}$; estimated matrices are denoted as $\bm{\widehat{X}}$; encrypted matrices are denoted as $\mathbb{X}$. We denote the $\ell_1$ norm of matrix as $\lVert \bm{X}\rVert_1 = \sum_{i,j}|\bm{X}(i,j)|$; and the Frobenius norm of a matrix is defined as $\lVert \bm{X}\rVert_F = \left( \sum_{i,j}\bm{X}(i,j)^2\right)^{1/2}$. `$\bm{X}\bm{Y}$' represents the matrix multiplication of two matrices, while `$\bm{X}\cdot \bm{Y}$' represents the element-wise multiplication. Then we give our definitions of matrices as following: 

$\emph{Environment Matrix}$ (EM): is an $n \times t$ matrix defined as $\bm{A}$. Considering multiple attributes, it is extended to $\bm{A}_k = (\bm{a}_{ij})_{n \times t}$, $k = 1,\,2,\,...,\,s$. $\bm{A}_k$ represents that the entire values of $s$ attributes from a set of $n$ nodes in $t$ time stamps are successfully gathered, i.e., ideally no data loss.

$\emph{Binary Index Matrix}$ (BIM): is an $n \times t$ matrix, which indicates whether the data points at the corresponding positions in the EM are missing. BIM is defined by B as:
$$
\bm{B}=(b_{ij})_{n \times t}=\left\{
\begin{array}{rcl}
0\quad &{\text{if}\;a_{ij}\;\text{is}\;\text{missing.}} \\ 
1\quad &{\text{otherwise.}}
\end{array}
\right.    
$$

$\emph{Sensed Matrix}$ (SM): is an $n \times t$ matrix, which records the sampled attributes collected from WSNs. Due to the phenomena of data missing, elements in SM are either $\bm{a}_{ij}$ measured by sensors or 0. Thereby, an SM is the incomplete EM, which is defined as: $\bm{S}=\bm{B}\cdot\bm{A}$.

$\emph{Recovered Matrix}$ (RM): is generated by interpolating the missing data in the SM to approximate the complete matrix EM. We use $\bm{\widehat{A}}$ to denote RM.

$\emph{Compressive Sensing}$ (CS): is an efficient recovery approach utilized to estimated missing values in SM. The CS procedure is denoted by $\emph{f}_{CS}$, thus, $\widehat{A}$ = $\emph{f}_{CS}(S)$.
\subsection{Privacy-Preserving Recovery Model}
\begin{figure}
    \centering
    \includegraphics[totalheight=3cm,width=\linewidth]{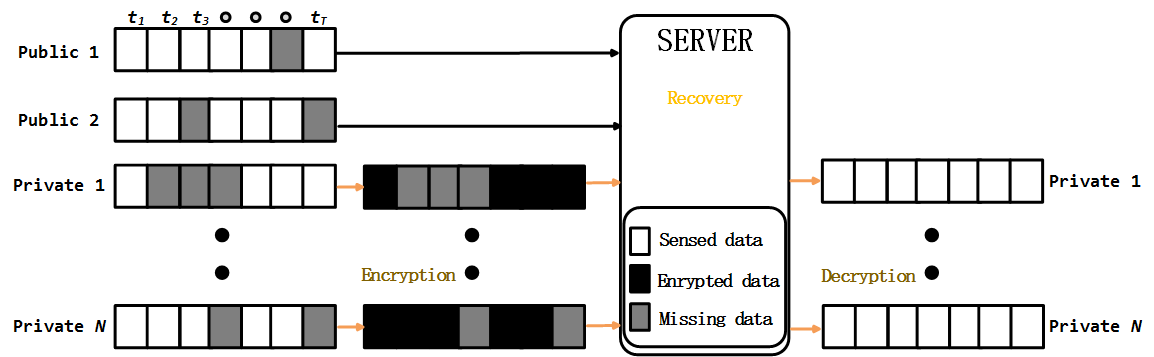}
    \caption{Privacy-preserving recovery model.}
    \label{fig:privacy}
\end{figure}
We consider a system consisting of two types of mobile users: public and private users. Public users are willing to share their data but private ones are the complete opposite. We illustrate the privacy-preserving recovery model as shown in Fig. 1. Then, we consider the accurate and privacy-preserving data recovery problem. The standard CS meets our requirements, but the objective of avoiding data exposure is unfortunately contrary to the basic requirement of CS data transmission. To address this dilemma, the PPCS scheme is proposed in the trajectory field. We elaborately apply PPCS to sensory data recovery. The procedures in detail is described in Section III.

\subsection{Multi-Attribute Assistance Model}
Analyzing the real research datasets of WSNs, we aim at discovering some features of common sensory data contributing to our data recovery approach. For instance, the datasets from IntelLab \cite{intel} and GreenOrbs \cite{liu2013does} are studied. The format of those data packet can be simplified as following:

\begin{tabular}{|c|c|c|c|c|} %l(left)居左显示 r(right)居右显示 c居中显示
\hline 
SensorID&Time&Attribute 1&Attribute 2&...\\
\hline  
\end{tabular}

Our target is to recover the complete series of these attributes matrices from their sampled subsets of entries which represent the incomplete data gathered by WSNs. We denote it as $\emph{Multi-Attribute Data Recovery}$ (MADR) problem. In this case, it is essential for us to mine the relations among multiple attributes. That means multiple matrices are evaluated conjointly. To simplify, we discuss the MADR problem in the condition of two attributes as an example. The basic analysis and evaluation of our approach can be easily extended into the condition of more attributes.
\subsection{Problem Statement/Formulation}
\subsubsection{Ground Truth}
After studying the original datasets gathered from GreenOrbs and IntelLab, we reveal that the data loss rates reach 35$\%$ and 23$\%$. In order to guarantee the integrity of ground truth, we perform preprocessing method on the raw data to filter and construct small but completed matrices as the "ground truth" matrices, which considers the maximization of the integrality in both time and space. From each dataset we extract subsets of two attributes: humidity and light.
\subsubsection{Low-rank Property}
As a matter of fact, there exists a strong correlation among the readings of neighboring sensors or time periods. Therefore, we can analyze the inherent structure or redundancy of datasets to ensure the low-rank property of matrices, which is the requirement for our fundamental CS operation.
We adopt the singular value decomposition (SVD). According to SVD theorem, an $n \times t$ matrix $\bm{X}$ has the decomposition as:
\begin{equation}
    \bm{X} = \bm{U}\bm{\Lambda}\bm{V}^{\top} = \sum_{i=1}^{\text{min}(n,t)}\sigma_i u_i v_i^{\top},
\end{equation}
where $\bm{U}$ and $\bm{V}$ are two unitary matrices with size $n\times n$ and $t\times t$. $(\cdot)^{\top}$ is the transpose operator, and $\bm{\Lambda}$ is an $n\times t$ diagonal matrix including all singular values $\sigma_i$ of $\bm{X}$. The singular values $\sigma_i$ are sorted as $\sigma_i \geq \sigma_{i+1}$, $i=1,\,...,\,min(n,t)$, where $min(n,t)$ is the number of singular values. The rank of the matrix $\bm{X}$, denoted by $r$, represents the number of its non-zero singular values. The low-rank property is stated when $r \ll min(n,t)$. Moreover, the matrix can be considered to be approximate low-rank, if the top-$\widehat{r}$ has a good approximation of the total singular values as: $\sum_{i=1}^{\widehat{r}} \sigma_{i} \approx \sum_{i=1}^{\text{min}(n,t)} \sigma_{i}$.
\subsubsection{Multi-Attribute Correlation} Here we consider solving MADR problem. Formally, when $s$ = 2, multi-attribute data recovery problem is defined as: Give SMs $\bm{S}_1$, $\bm{S}_2$, find the solution as $\widehat{\bm{A}_1}$ and $\widehat{\bm{A}_2}$, i.e.,
\begin{equation}
\begin{aligned}
    \text{minimize:} \quad &\lVert \widehat{\bm{A}_1} - \bm{A}_1\rVert_F + \mu\lVert \widehat{\bm{A}_2} - \bm{A_2}\rVert_F,\\
    \text{subject to:} \quad &\bm{B}_1 \cdot \bm{\widehat{A}}_1=\bm{B}_1 \cdot \bm{A}_1,\\
    &\bm{B}_2 \cdot \bm{\widehat{A}}_2=\bm{B}_2 \cdot \bm{A}_2,
\end{aligned}
\end{equation}
where $\lVert \cdot\rVert_F$ expresses the Frobenius norm. Here we add parameter $\mu$ as a tradeoff coefficient in order to avoid the overshadow problem between two matrices, because of the inequality in the magnitudes of attributes. 

$\textbf{Joint Sparse Decomposition}$ (JSD) \cite{chen2013multiple} is the approach to jointly dividing multi-attribute matrices into a public component of matrix $\bm{U}$ and two private components of matrices $\bm{\Delta}_1$, $\bm{\Delta}_2$. Suppose two attributes $\bm{A}_1=(a_1^{(i)},\,...,\,a_1^{(i)})$ and $\bm{A}_2=(a_2^{(i)},\,...,\,a_2^{(i)})$. For both column vector $a_1^{(i)}$ and $a_2^{(i)}$, the goal is to split them into:
\begin{equation}
\begin{aligned}
    a_1^{(i)} &= u^{(i)} + \delta_1^{(i)},\\
    a_2^{(i)} &= u^{(i)} + \delta_2^{(i)},\\
    u^{(i)} &= \Psi v^{(i)},
\end{aligned}
\end{equation}
where $u^{(i)}$ is the public component, which is the multiplication of a wavelet basis $\Psi$ and a sparse vector $v^{(i)}$. The private components are respectively represented by $\delta_1^{(i)}$ and $\delta_2^{(i)}$. Considering the CS theory, $(v^{(i)},\delta_1^{(i)},\delta_2^{(i)})$ are obtained by solving an $\ell_1$-norm minimization problem as following:
\begin{equation}
    \widehat{\theta} = \text{argmin} \lVert \theta\rVert_1, s.t.\;a = \bm{H}\theta.
\end{equation}
where $\theta = (v^{(i)\top},\delta_1^{(i)\top},\delta_2^{(i)\top})^{\top}$, $a = (a_1^{(i)\top},a_2^{(i)\top})^{\top}$ and $\bm{H} = (\bm{\Psi},\bm{I},\bm{0};\bm{\Psi},\bm{0},\bm{I})$. Then $u^{(i)\top},\delta_1^{(i)\top}$ and $\delta_2^{(i)\top}$ are calculated from $\theta$.
Utilizing JSD to every column vector, $\bm{A}_1$ and $\bm{A}_2$ are decomposed as:
\begin{equation}
    \begin{aligned}
        \bm{A}_1 &= \bm{U} + \bm{\Delta}_1,\\
        \bm{A}_2 &= \bm{U} + \bm{\Delta}_2.
    \end{aligned}
\end{equation}
%%%%%%%%%%%%%%%%%%%%%%%%%%%%%%%%%%%%%%%
\section{Solutions}
\subsection{PPCS Approach}
To conquer the challenge of privacy issue in classic CS, we introduce the simple and feasible sensor data recovery approach $\emph{Privacy-Preserving Compressive Sensing}$ (PPCS) in this section. The suggested approach consists of three steps. Generally, they can be sequentially executed as encryption, recovery and decryption. 
\subsubsection{Encryption}
Define $f_{en}(\cdot)$ as the encryption function. With a sensed matrix $\bm{S}$, the encrypted matrix is denoted as $\mathbb{S} = f_{en}(\bm{S}),$. The encryptor in detail is stated as $\emph{K-Vector Perturbation}$(KVP), which is described as follows:

Firstly, K public vectors $D_1$, $D_2$, ..., $D_K$ are randomly downloads from all public vectors at the server or just randomly initialized, which is used to construct the encrypted $i$-th row vector $\mathbb{S}_i$. Then, a length-$(K+1)$ random vector $<\psi_{i,0},\psi_{i,1},\,...,\,\psi_{i,K}>$ is generated as a private key. Any key satisfies $\psi_{i,j}\in(0,1)$. The encryption operator is represented with the public vectors and the private key as:
\begin{equation}
    \mathbb{S}_i = (\psi_{i,0}\bm{S}_i + \psi_{i,1}D_1 + \,\cdots\, + \psi_{i,K}D_K)\cdot \bm{B}_i.
\end{equation}

Intuitively, the length of private key dominates the difficulty of decryption and the value of $K$ decides the privacy preservation strength. 

\subsubsection{Recover the encrypted data}
After gathering the encrypted components, the encrypted matrix $\mathbb{S}$ is an $n \times t$ matrix. Then, for recovering, CS operator $f_{cs}(\cdot)$ is applied to $\mathbb{S}$ and the recovery matrix is perceived as $\mathbb{\widehat{A}}$.
Simply, we adopt the typical CS recovery scheme, which is processed as follows:

$\mathbb{\widehat{A}}$ is separated into $\bm{L}$ and $\bm{R}$ partial matrices by the SVD-like factorization.
\begin{equation}
    \mathbb{\widehat{A}} = \bm{U}\bm{\Lambda}\bm{V}^{\top} = \bm{L}\bm{R}^{\top},
\end{equation}
where $\bm{L}=\bm{U}\bm{\Lambda}^{1/2}$, $\bm{R}=\bm{V}\bm{\Lambda}^{1/2}$. Then $\bm{L}$ and $\bm{R}$ matrices are estimated by
\begin{equation}
    \text{min}(\lVert \bm{B} \cdot (\bm{L}\bm{R}^{\top}) - \mathbb{S}\rVert_F^2+\lambda(\lVert \bm{L}\rVert_F^2 + \lVert \bm{R}^{\top}\rVert_F^2)),
\end{equation}
where the Lagrange parameter $\lambda$ serves as a tunable tradeoff between rank approximation and accuracy fitness. The solution of $\bm{L}$ and $\bm{R}$ can be attained by iterative computing. To summarize, the operator $f_{cs}$ is equal to solve $\mathbb{\widehat{A}} = \bm{L}\bm{R}^{\top}$ by $\bm{L}$ and $\bm{R}$ obtained from Eqn.(8) with input $\mathbb{S}$.

\subsubsection{Decryption}
When the recovery procedure is done, the recovered matrix denoted as  $\mathbb{\widehat{A}}$ can be delivered to local decryption operation as $f_{de}(\cdot)$. Also, the operation obtains the decrypted and estimated matrix $\bm{\widehat{A}}$ as 
\begin{equation}
    \bm{\widehat{A}}_i = f_{de}(\mathbb{\widehat{A}}_i) = (\mathbb{\widehat{A}}_i - (\psi_{i,1}D_1 +\,\cdots\,+ \psi_{i,K}D_K)\,/\,\psi_{i,0}).
\end{equation}

Finally $\bm{\widehat{A}}$ can be obtained by the union of every $i$-th row vector as the matrix $\bm{\widehat{A}}_i$. Considering the local decryption and the private key, the privacy of $\bm{\widehat{A}}$ is safely preserved.
In addition, $\psi_{i,0}$ is a special component in private key. As $f_{de}$ operation in Eqn.(9) indicates, $\psi_{i,0}$ determines the weight of the original vectors in the encrypted vector. For one thing, $\psi_{i,0}$ cannot be too small, because under this condition, the weight of $\bm{\widehat{A}}_i$ hidden in the encrypted $\mathbb{\widehat{A}}_i$ is too small, which leads to a poor recovery accuracy. For another, $\psi_{i,0}$ cannot be too large. With $\psi_{i,0}$ close to 1, $\bm{\widehat{A}}_i = \mathbb{\widehat{A}}_i$, which losses the strength of privacy preservation. Empirically, $\psi_{i,0}$ is set in the range of $[0.2,0.8]$.

\subsection{Multi-Attribute Assistance}
\subsubsection{Normalization}
In order to tackle the issue of the matrix overshadowing problem, the parameter $\mu$ is supplemented in Eqn.(2). However, it is difficult to find the best value of $\mu$ due to the indistinct relationship between $\bm{A}_1$ and $\bm{A}_2$. So we simply normalize each matrix and set $\mu$ = 1. The real maximum value is possible to loss, hence we adopt the maximum value in gathered datasets instead.

\subsubsection{Low-Rank Matrix Approximation}

The problem in Eqn.(2), restricted by variables in two matrices, is difficult to solve in closure form. Since the inherited low-rank features are stated in Section II, we convert this problem to a rank minimization problem as:
\begin{equation}
    \text{min}(rank(\bm{\widehat{A}}_i)),\quad \text{s.t.}\;\bm{S}_i = \bm{B}_i \cdot \bm{A}_i.
\end{equation}

However, the rank calculation $rank(\cdot)$ is not convex. We adopt SVD-like factorization as Eqn.(7). Thus $\text{min}(rank(\bm{\widehat{A}}))$ is also solvable by figuring $\bm{L}$ and $\bm{R}$, which is redefined as
\begin{equation}
    \text{min}(\lVert \bm{L}\rVert_F^2 + \lVert \bm{R}^{\top}\rVert_F^2).
\end{equation}
 
\subsubsection{Compressive Sensing-based Joint Matrix Decomposition}
There is still a problem against us that no connection is established between $\bm{A}_1$ and $\bm{A}_2$ in Eqn.(11). To exploit the inherent correlation, we apply JSD to our CS operation. Thus, the approximation $\bm{\widehat{A}}_1$ and $\bm{\widehat{A}}_2$ are separated by JSD as Eqn.(5).
Since $\bm{\widehat{U}}$, $\bm{\widehat{\Delta}}_1$, $\bm{\widehat{\Delta}}_2$ inherit the low-rank feature, the problem is reformulated as:
\begin{equation}
    \begin{aligned}
        \text{minimize:}\quad  &\lVert \bm{\widehat{U}}\rVert_* + \lVert \bm{\widehat{\Delta}}_1\rVert_* + \lVert \bm{\widehat{\Delta}}_2\rVert_*,\\
        \text{subject to:}\quad  &\bm{B}_1 \cdot (\bm{\widehat{U}}+\bm{\widehat{\Delta}}_1) = \bm{B}_1 \cdot \bm{A}_1,\\
        &\bm{B}_2 \cdot (\bm{\widehat{U}}+\bm{\widehat{\Delta}}_2) = \bm{B}_2 \cdot \bm{A}_2.
    \end{aligned}
\end{equation}

Moreover, we use SVD-like factorization to minimize objective in Eqn.(12), which is rewritten as:
\begin{equation}
    \sum_L \lVert \bm{L}_j\rVert_F^2 + \sum_R \lVert \bm{R}_j\rVert_F^2,\quad j = 1,2,U.
\end{equation}
where $\bm{L}_U$, $\bm{L}_1$, $\bm{L}_2$ are $n \times r$ matrices and $\bm{R}_U$, $\bm{R}_1$, $\bm{R}_2$ are
$t \times r$ matrices. $ \bm{\widehat{U}} = \bm{L}_U \bm{R}_U^{\top}, \bm{\widehat{\Delta}}_1 = \bm{L}_1 \bm{R}_1^{\top}$ and $\bm{\widehat{\Delta}}_2 = \bm{L}_2 \bm{R}_2^{\top}$. To avoid overfitting, we adopt the Lagrange multiplier method which converts the problem to a non-stationary optimization problem, i.e.,
\begin{equation}
    \begin{aligned}
        \text{minimize:}\quad &\lVert \bm{B}_1 \cdot (\bm{L}_U \bm{R}_U^{\top} + \bm{L}_1 \bm{R}_1^{\top}) - \bm{S}_1\rVert_F^2\\
        + &\lVert \bm{B}_2 \cdot (\bm{L}_U \bm{R}_U^{\top} + \bm{L}_2 \bm{R}_2^{\top}) - \bm{S}_2\rVert_F^2\\
        + &\lambda(\sum_L \lVert \bm{L}_j\rVert_F^2 + \sum_R \lVert \bm{R}_j\rVert_F^2).
    \end{aligned}
\end{equation}

This equation is the core of MAA component, which can be solved because 1) $\bm{B}_1$, $\bm{B}_2$, $\bm{S}_1$, $\bm{S}_2$ are known; 2) each square of the Frobenius norm is non-negative; 3) the optimal value can be reached by minimizing all non-negative parts to zero. Hence, $\bm{\widehat{A}}_1$ and $\bm{\widehat{A}}_2$ can be estimated by Eqn.(14) and Eqn.(5). Combining Eqn.(14) with Eqn.(8), we propose our novel approach as following:\\
\hspace*{0.12in} {\bf PPCS-MAA} is to recover encoded multiple (two as example) attributes-based sensory matrices according to:
\begin{equation}
    \begin{aligned}
        \text{minimize:}\quad &\lVert \bm{B}_1 \cdot (\bm{L}_U \bm{R}_U^{\top} + \bm{L}_1 \bm{R}_1^{\top}) - \mathbb{S}_1\rVert_F^2\\
        + &\lVert \bm{B}_2 \cdot (\bm{L}_U \bm{R}_U^{\top} + \bm{L}_2 \bm{R}_2^{\top}) - \mathbb{S}_2\rVert_F^2\\
        + &\lambda(\sum_L \lVert \bm{L}_j\rVert_F^2 + \sum_R \lVert \bm{R}_j\rVert_F^2).
    \end{aligned}
\end{equation}
\hspace*{0.12in} {\bf Extension:} The MAA component suits the condition of more attributes as well. For example, if we measure k attributes in one WSN, denoted as $\bm{A}_1$, $\bm{A}_2$, ..., $\bm{A}_k$, the minimization objective in Eqn.(12) is rewritten as following:
\begin{equation}
    \lVert \bm{\widehat{U}}\rVert_* + \lVert \bm{\widehat{\Delta}}_1\rVert_* + \lVert \bm{\widehat{\Delta}}_2\rVert_* +\, \cdots +\, \lVert \bm{\widehat{\Delta}}_k\rVert_*.
\end{equation}

This equation can be solved by the similar method of the two-attribute case above, which is concretely shown in Alg.1.

\subsection{Algorithm}
\begin{algorithm}
\caption{PPCS-MAA}
{\bf Input:} SMs $\bm{S}_1$, $\bm{S}_2$ with BIMs $\bm{B}_1$, $\bm{B}_2$, parameters $\lambda, r, k$.\\
{\bf Main Procedure:}
\begin{algorithmic}[1]
\State Initialize the public and private key $\bm{D}, \bm{\Psi}$.
\State Normalize the observed matrices $\bm{S}_1, \bm{S}_2$.
\State \textbf{Encrypt} $\mathbb{S}_1 \leftarrow [\bm{S}_1\,\bm{D}]\bm{\Psi};\quad \mathbb{S}_2 \leftarrow [\bm{S}_2\,\bm{D}]\bm{\Psi}$. 
\State Initialize random $\bm{L}_U, \bm{L}_1, \bm{L}_2, \bm{R}_1, \bm{R}_2$.
\State Objective $y \leftarrow Eqn.(15)$.
\While{$y$ is not converged}
    \State $\bm{X}_1 \leftarrow \mathbb{S}_1 - \bm{B}_1 \bm\cdot (\bm{L}_1 \bm{R}_1^{\top}); \bm{X}_2 \leftarrow \mathbb{S}_2 - \bm{B}_2 \bm\cdot (\bm{L}_2 \bm{R}_2^{\top})$;
    \State $\bm{\widehat{R}}_U \leftarrow \text{crossInverse}(\bm{B}_1,\bm{B}_2,\bm{L}_U,\lambda,r,\bm{X}_1,\bm{X}_2)$
    \State $\bm{\widehat{L}}_U \leftarrow \text{crossInverse}(\bm{B}_1^{\top},\bm{B}_2^{\top},\bm{R}_U,\lambda,r,\bm{X}_1^{\top},\bm{X}_2^{\top})$
    \State $\bm{Z}_1 \leftarrow \mathbb{S}_1 - \bm{B}_1 \bm\cdot (\bm{L}_U \bm{R}_U^{\top}); \bm{Z}_2 \leftarrow \mathbb{S}_2 - \bm{B}_2 \bm\cdot (\bm{L}_U \bm{R}_U^{\top})$;
    \State $\bm{\widehat{R}}_1 \leftarrow \text{singleInverse}(\bm{B}_1,\bm{L}_1,\lambda,r,\bm{Z}_1)$
    \State $\bm{\widehat{L}}_1 \leftarrow \text{singleInverse}(\bm{B}_1^{\top},\bm{R}_1,\lambda,r,\bm{Z}_1^{\top})$
    \State $\bm{\widehat{R}}_2 \leftarrow \text{singleInverse}(\bm{B}_2,\bm{L}_2,\lambda,r,\bm{Z}_2)$
    \State $\bm{\widehat{L}}_2 \leftarrow \text{singleInverse}(\bm{B}_2^{\top},\bm{R}_2,\lambda,r,\bm{Z}_2^{\top})$
\EndWhile
\State \textbf{end while}
\State $\mathbb{\widehat{A}}_1 \leftarrow \bm{\widehat{L}}_U \bm{\widehat{R}}_U^{\top} + \bm{\widehat{L}}_1 \bm{\widehat{R}}_1^{\top}$; $\mathbb{\widehat{A}}_2 \leftarrow \bm{\widehat{L}}_U \bm{\widehat{R}}_U^{\top} + \bm{\widehat{L}}_2 \bm{\widehat{R}}_2^{\top}$;
\State \textbf{Decrypt} $\bm{\widehat{A}}_1 \leftarrow \mathbb{\widehat{A}}_1-\bm{D}\bm{\Psi}\,/\,\psi_{i,0}; \bm{\widehat{A}}_2 \leftarrow \mathbb{\widehat{A}}_2-\bm{D}\bm{\Psi}\,/\,\psi_{i,0}$;
\end{algorithmic}
{\bf Output:}  $\bm{\widehat{A}}_1 \leftarrow \alpha_1 \bm{\widehat{A}}_1; \bm{\widehat{A}}_2 \leftarrow \alpha_2 \bm{\widehat{A}}_2$;

{\bf Procedure} {$\bm{Y} = \text{singleInverse}(\bm{B},\bm{L},\lambda,r,\bm{X})$:}
\begin{algorithmic}[1]
\For{$i=1$ to $t$}
    \State $\bm{P}_i \leftarrow [Diag(\bm{B}(:,i))\bm{L}; \sqrt{\lambda}\bm{I}_r]$; $\bm{Q}_i \leftarrow [\bm{X}(:,i);\bm{0}_r]$
    \State $\bm{Y}(:,i) = (\bm{P}_i^{\top} \bm{P}_i)\backslash (\bm{P}_i^{\top} \bm{Q}_i)$
\EndFor
\end{algorithmic}
{\bf Procedure} {$\bm{Y} = \text{crossInverse}(\bm{B}_1,\bm{B}_2,\bm{L},\lambda,r,\bm{X}_1,\bm{X}_2)$:}
\begin{algorithmic}[1]
\For{$i=1$ to $t$}
    \State $\bm{P}_i \leftarrow [Diag(\bm{B}_1(:,i))\bm{L};Diag(\bm{B}_2(:,i))\bm{L}; \sqrt{\lambda}\bm{I}_r]$;
    \State $\bm{Q}_i \leftarrow [\bm{X}_1(:,i);\bm{X}_2(:,i);\bm{0}_r]$
    \State $\bm{Y}(:,i) = (\bm{P}_i^{\top} \bm{P}_i)\backslash (\bm{P}_i^{\top} \bm{Q}_i)$
\EndFor
\end{algorithmic}
\end{algorithm}
 Firstly, we do initialization and preprocess the dataset with sampling the EMs by BMs respectively. Secondly, we generate the vectors of KVP approach to perform the $f_{en}(\cdot)$ operation. Thirdly, we present the core calculations of MAA component to solve Eqn.(15) here. Fourthly, we obtain the RMs and execute $f_{de}(\cdot)$ to complete the output. The algorithm solves the problem in an iterative manner. In the Beginning, all $\bm{L}$ and $\bm{R}$ matrices are initialized randomly except $\bm{R}_U$. Then, with $\bm{L}_U$ fixed, $\bm{R}_U$ can be calculated from the initialized matrices by solving the equation:
 \begin{equation}
     \left[
     \begin{array}{rcl}
        \bm{B}_1 \cdot(\bm{L}_U\bm{R}_U^{\top})  \\
        \bm{B}_2 \cdot(\bm{L}_U\bm{R}_U^{\top})  \\
            \sqrt{\lambda}\bm{R}_U^{\top}
     \end{array}\right] = 
     \left[
     \begin{array}{rcl}
        \mathbb{S}_1 \cdot(\bm{L}_1\bm{R}_1^{\top})  \\
        \mathbb{S}_2 \cdot(\bm{L}_2\bm{R}_2^{\top})  \\
            \bm{0}
     \end{array}\right].
 \end{equation}
 
Eqn.(17) can be treated as a linear least squares problem. The inverse procedure $crossInverse$ computes $\bm{R}_U$ and $\bm{L}_U$. The other $\bm{L}$ and $\bm{R}$ matrices are obtained by similar inverse procedure $singleInverse$. Furthermore, in Alg.1, the rank approximation $r$ and the lagrange tradeoff coefficient $\lambda$ are influential in the accuracy of estimation. The key operation in Alg.1 is the encryption, CS recovery with inverse computation, and decryption. The KVP operation processes $k+1$ vectors with size $1\times t$ in encryption and decryption procedures, which both require complexity of $O((k+1)t)$. The more complicate task is CS computing, whose convergence to optimal requires the complexity of $O(nrt)$. Together, the algorithm requires the computational complexity of $O(knrtm)$, where $m$ is the iteration times.
% We obtain the singular values from GreenOrbs and IntelLab. 20$\%$ top singular values contribute to over 90$\%$ energy in the datasets. Hence, both datasets are specified. The approximate low-rank structure and the standard in CS operation is satisfied to perform a promising recovery accuracy. Thus our evaluation uses $r = 20\% min(n, t)$. 
%%%%%%%%%%%%%%%%%%%%%%%%%%%%%%%%%%%%%%%

\begin{figure}
    \centering
    \includegraphics[totalheight=4.75cm,width=9.5cm]{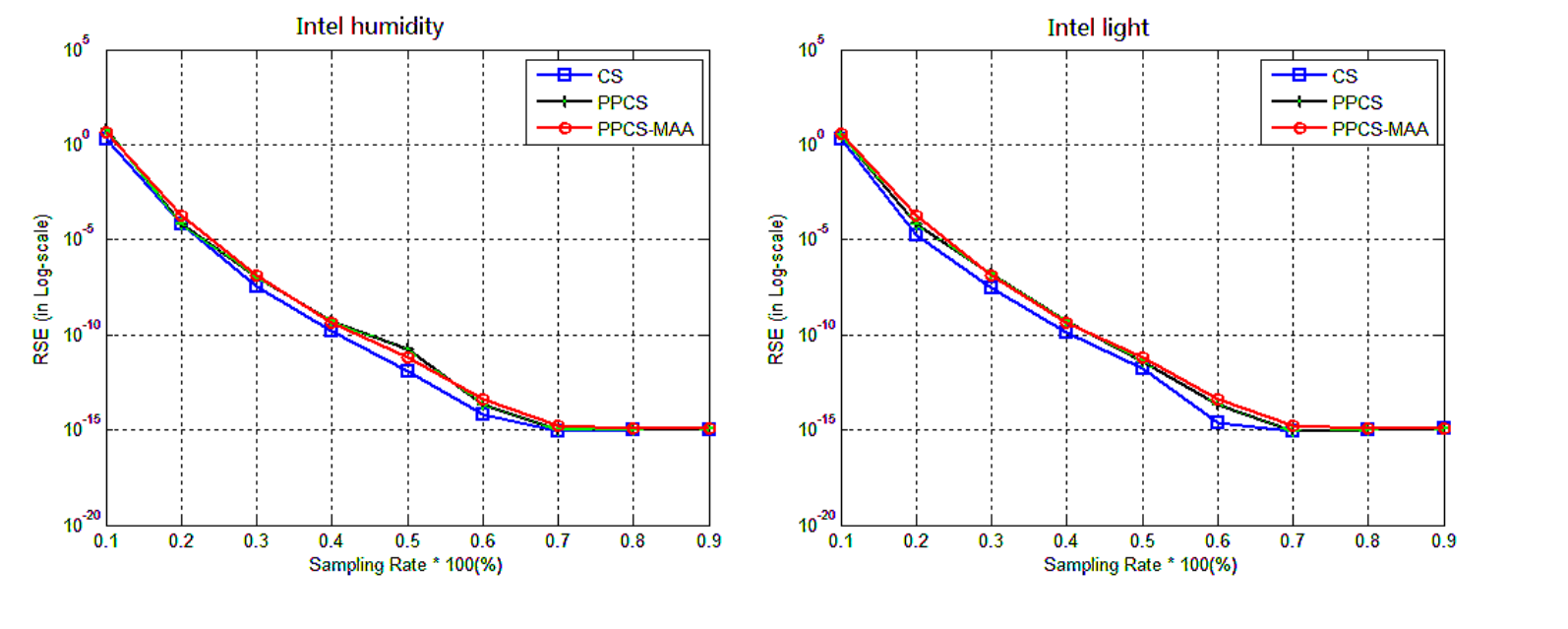}
    \caption{Comparisons for humidity, light of the IntelLab data.}
    \label{fig:intel}
\end{figure}

\begin{figure}
    \centering
    \includegraphics[totalheight=4.5cm,width=9.1cm]{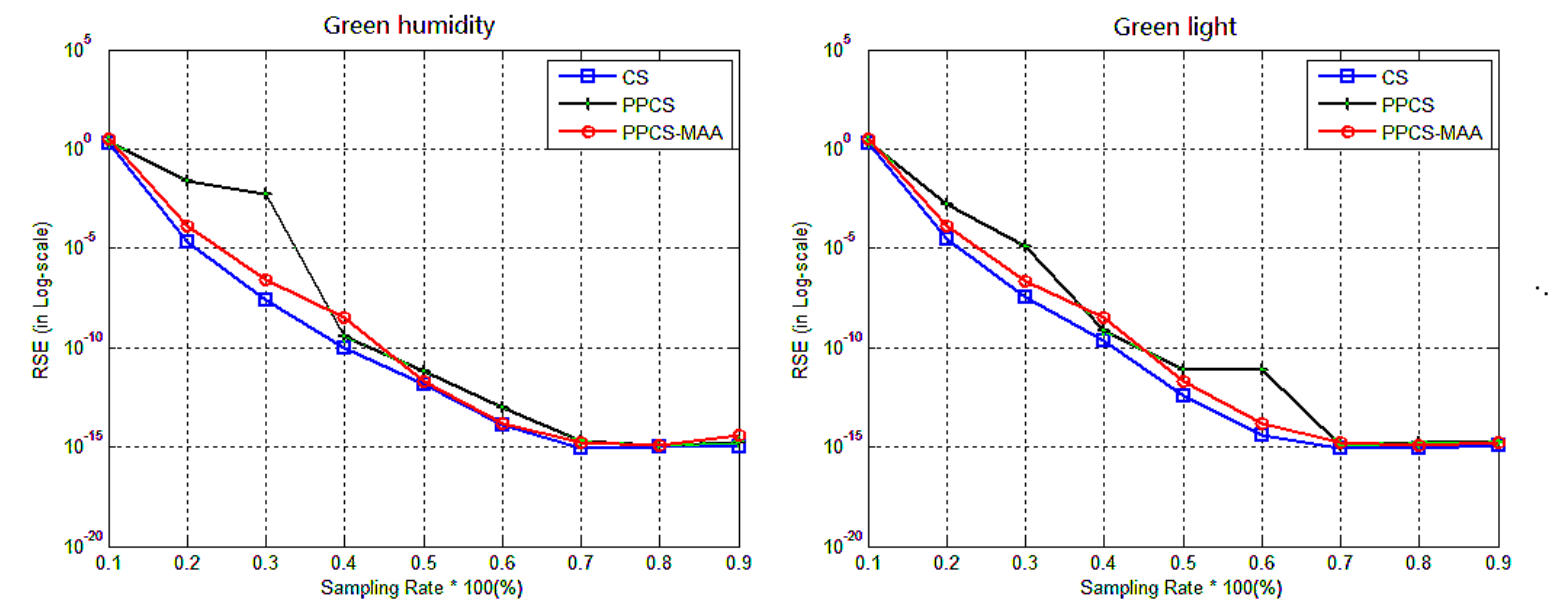}
    \caption{Comparisons for humidity, light of the GreenOrbs data.}
    \label{fig:greenorbs}
\end{figure}

\section{Performance Evaluations}

\subsection{Evaluation Methodology}
{\bf Ground Truth:} Our datasets are from the IntelLab and GreenOrbs projects. Two types of attributes are chosen in each project: Indoor-humidity, Indoor-light, GreenOrbs-humidity, GreenOrbs-light. We preprocess the original data in Section II-D(1) to extract ``ground truth" matrices.

{\bf Compared Methods:} To verify the effectiveness of our approach, two methods are chosen for comparison. They are the classic method CS \cite{liu2017ls} and the state-of-the-art method, PPCS \cite{kong2015privacy}.

{\bf Metric:} We measure the normalized square error (NSE) defined as follows:
\begin{equation}
    \text{NSE} = \frac{\lVert \bm{A}-\bm{\widehat{A}}\rVert_F}{\lVert \bm{A}\rVert_F}.
\end{equation}

\subsection{Performance Results on Real Sensory Datasets}
We randomly drop entries with missing data rate $10\%$, $20\%$, ..., $90\%$, i.e., with sampling rate $90\%$, $80\%$, ..., $10\%$. After preprocessing the remaining data, i.e., outlier processing, we apply the above four schemes for data recovery. Then we conduct 15 iterations for each case and record the average result over these 10 runs. From Figs. 2 and 3, we find that as the sampling rate increases (or the data loss rate decreases), the recovery errors for all the three methods decrease. PPCS-MAA performs much better than PPCS, because PPCS does not consider multiple attributes correlation. The performance of PPCS-MAA is close to CS, which means our method satisfies the demand of high accuracy in the condition of good privacy. Generally, our scheme achieves recovery error $\leq 1\%$ for sampling rate $\geq 20\%$ and almost exact recovery for sampling rate $\geq 40\%$. Note that although PPCS performs as well as our method in terms of privacy, its accuracy catches up with our method only for sampling rate $\geq 40\%$ due to the absence of MAA. Fig. 2 shows the recovery performance for the IntelLab data.
For sampling rate $20\%, 50\%, 60\%$, and $70\%$, PPCS is worse than PPCS-MAA. The recovery errors of PPCS and PPCS-MAA decrease further with more samples, keeping pace with the standard CS. They all converge to nearly zero recovery error for sampling rate $\geq 70\%$.
In Fig. 3, the results turn out to be similar for GreenOrbs, except that PPCS performs a little unsteadily. Comparing GreenOrbs with IntelLab data, we observe bigger gaps between PPCS and PPCS-MAA/CS.  

\section{Conclusion}
In this paper, we have investigated the data privacy issue in the presence of sensory data recovery with multiple attributes. After we discussed the drawback of existing works, an effective privacy-preserving recovery service was modeled. From the datasets in WSNs, we mined the low-rank feature and multi-attribute correlation. Driven by these observations, a PPCS-MAA approach was proposed to accurately recover the missing data, which included the privacy preservation to compressive sensing with multi-attribute assistance. Given the theoretical results of analysis of accuracy and complexity, our scheme is as accurate as standard CS completion methods. The proposed algorithm not only effectively preserves the data privacy with satisfying recovery accuracy, but also combines multiple attributes recovery system which will help to control cost or data trading. Data-driven simulations illustrate that the approach is expected to have wide real-world applications.

\bibliographystyle{IEEEbib}
\bibliography{reference}
\end{document}